# Generalization of Gamow states to multi-particle decay

Jonathan F. Schonfeld
Center for Astrophysics | Harvard and Smithsonian
60 Garden St., Cambridge, Massachusetts 02138 USA
jschonfeld@cfa.harvard.edu
ORCID ID# 0000-0002-8909-2401

**Abstract** For single-particle nonrelativistic quantum mechanics, a Gamow state is an eigenfunction of the Hamiltonian with complex eigenvalue. Gamow states are not normalizable; they depend on time via the usual multiplier exp(-i$Et$) supplemented by a cutoff at an expanding wavefront. Gamow states have been used to extend nuclear shell models: they are to metastable states as normalizable eigenfunctions are to bound states. In this paper we generalize Gamow states to decays with multiple outgoing particles. We derive the exact form of the expanding wavefront, even for relativistic outgoing particles. In the non-relativistic limit we derive the exact form of the multi-particle Gamow state contribution to the system propagator (Green's function).

**Keywords** Gamow state; Schroedinger equation; wavefront; decay; asymptotic behavior

## 1. Introduction

This paper derives new results about the asymptotic behavior of wavefunctions that arise from the quantum-mechanical decay of metastable states. The results are elementary but previously unknown, and enlarge our intuition about what quantum mechanical wavefronts look like in the real world.

For a single spinless nonrelativistic particle of energy $E$ and mass $m$ interacting in three dimensions with a localized, spherically symmetric potential, the asymptotic situation is already very familiar: For a bound state ($E<0$), the wavefunction at time $t$ and large radius $r$ is proportional to a spherical harmonic multiplied by

$$\frac{1}{r} exp\left(-r\sqrt{2m|E|}\right) exp(-iEt), \qquad (1)$$

($\hbar$ set to unity). For a continuum state ($E>0$), this large-radius multiplier becomes

$$\frac{1}{r} exp(\pm ir\sqrt{2mE}) exp(-iEt). \qquad (2)$$

Finally, for a narrow resonance at $E=E_0 - i\Gamma/2$ embedded in the continuum, the large-radius multiplier becomes [1,2]

$$\sim \frac{1}{r} \exp\left( ir \sqrt{\frac{mE_0}{2}} \right) \exp\left[ -i\left( E_0 - \frac{i\Gamma}{2} \right)\left( t - \frac{r}{v_0} \right) \right] \theta\left( t - \frac{r}{v_0} \right), \qquad (3)$$

where speed $v_0 = (2mE_0)^{1/2}$. The cutoff in Equation (3) defines the leading edge or wavefront of the particle emitted in the decay of the resonance. After restoring the spherical harmonic, what multiplies the step function in Eq. (3) is naively an eigenfunction $u$ of the Hamiltonian with complex eigenvalue $E_0 - i\Gamma/2$. This eigenfunction grows exponentially at large distance and so does not have a conventional square norm. Nevertheless, its contribution to the space-energy Green's function (or propagator) $G(\mathbf{x},\mathbf{x}',E)$ for outgoing waves (see below) is closely analogous to that of a discrete bound state [3]

$$\frac{u(\mathbf{x})u(\mathbf{x}')}{\mathcal{N}\left( E - E_0 + \frac{i\Gamma}{2} \right)}, \qquad (4)$$

where the pseudo-norm is given by

$$\mathcal{N} = \int_{r<R} u^2(\mathbf{x}) d^3x - \frac{i}{2k_D} \oint_{r=R} u^2(\mathbf{x}) d^2x, \qquad (5)$$

in the limit of very large radius $R$. In Equation (5), $k_D$ is defined by

$$k_D \equiv \sqrt{2m\left( E_0 - \frac{i\Gamma}{2} \right)}. \qquad (6)$$

The exponentially growing function $u$ is called a Gamow state. Our specific aim in this paper is to generalize Equations (3) and (5) to situations in which the resonance decays into *multiple* particles of different masses. Equation (3), but not (5), will also generalize to multiple *relativistic* particles.

The idea of the Gamow state is important for several reasons.

- It simplifies quantum mechanical intuition by providing a unified formalism that treats a metastable state as just another kind of bound state [4].
    o This has led to an enlargement of nuclear models to include metastable single-particle shells [5].
- It provides a way to think about quantum mechanical wavefronts in complex circumstances. An understanding of wavefronts is essential to fully understanding the physical origin of canonical quantum measurement behavior [6].

- It has been proposed [7] that a Gamow-state-to-continuum transition explains the Mott problem regarding the appearance of a collimated cloud chamber track from a spherically symmetric radioactive decay. This has been examined experimentally [8].

The remainder of this paper is organized as follows. In the next section, we establish basic formal structure for the derivations that follow. In Section 3, we generalize the wavefront in Eq. (3) to multi-particle final states. In Section 4, we generalize the Gamow-state pseudo-norm to multi-particle final states. Section 5 contains concluding remarks.

## 2. General multi-particle final state

Consider a final state of $N$ particles with masses $m_n$, momenta $p_n$, position coordinates $\mathbf{x}_n$ and radial coordinates $r_n$. Outside the interaction region, the particles are governed by free dynamics: The $N$-particle wavefunction ($N+1$-point function in the relativistic field-theoretic case) is a sum/integral of products of $N$ factors, one factor for each $n$. Each factor is the product of a spherical harmonic (including indices for spin) in the solid angle of $\mathbf{x}_n$ and an outgoing spherical Bessel function of $p_n r_n$. The overall sum is over the angular momentum and spin indices of all the particles, and the integral is over the multiplet $\mathbf{p}$ of outgoing momenta. The overall integral is constrained by the condition $E_F(\mathbf{p})=E$, where $E$ is the particular overall energy of the decay and the function $E_F$ indicates how energy depends on momenta. In the nonrelativistic case,

$$E_F(\mathbf{p}) \equiv \sum_n \frac{p_n^2}{2m_n}. \tag{7}$$

In the relativistic case ($c$ set to unity),

$$E_F(\mathbf{p}) \equiv \sum_n \sqrt{m_n^2 + p_n^2}. \tag{8}$$

To promote clarity through maximum generality, we shall avoid specifying a particular functional form for $E_F$ unless absolutely necessary.

At large distances, the spherical Bessel functions are proportional to $(p_n r_n)^{-1}\exp(ip_n r_n)$, so that the overall wavefunction is a sum/integral of products of spherical/spin harmonics multiplied by a function of $\mathbf{p}$ and $\mathbf{r}$ that varies slowly ($\mathbf{r}$ is the multiplet with components $r_n$), all multiplied by

$$\exp(i\mathbf{p} \cdot \mathbf{r}). \tag{9}$$

The derivations that follow focus on the object in Equation (9).

## 3. Wavefront (leading edge) of multi-particle final state

For large $\mathbf{r}$, the integral of Equation (9) over $\mathbf{p}$ is dominated by a stationary phase point $\mathbf{p}_s$, such that for all *relevant* deviations $\delta\mathbf{p}$ around $\mathbf{p}_s$, we have

$$\delta \boldsymbol{p} \cdot \boldsymbol{r} = 0. \tag{10}$$

But the constant-energy constraint $E_F = E$ defines "all relevant deviations" via

$$\sum_n \delta p_n \frac{\partial E_F}{\partial p_n} = 0. \tag{11}$$

It follows that, in an obvious notation,

$$\boldsymbol{r} = \tau \boldsymbol{\nabla}_p E_F \tag{12}$$

at the stationary phase point, where $\tau$ is a scalar with dimensions of time. Since $\boldsymbol{\nabla}_p E_F$ is the canonical representation of classical velocity **v** at momentum **p**, we can equivalently write

$$\boldsymbol{v} = \frac{\boldsymbol{r}}{\tau} \tag{13}$$

and consider $E_F$ as a function of **v** instead of **p**. Accordingly, $\tau$ is an implicit function of **r** and $E$, defined by the condition

$$E_F\left(\boldsymbol{v} = \frac{\boldsymbol{r}}{\tau}\right) = E. \tag{14}$$

In words, if the $N$ particles all travel at constant (but different) speeds from the origin to their various destinations $r_n$ in the same time, then $\tau$ is the travel time that forces all those constant speeds to constitute total kinetic energy $E$.

If the resonance at $E = E_0 - i\Gamma/2$ is narrow, then we obtain the leading edge by linearizing $\boldsymbol{p}_s \cdot \boldsymbol{r}$ about $E = E_0$ and then integrating

$$e^{-i(E_0 + \delta E)t} e^{[i\boldsymbol{p}_s \cdot \boldsymbol{r}]_{E=E_0}} \frac{\exp\left(i\delta E \left[\frac{\partial}{\partial E} \boldsymbol{p}_s \cdot \boldsymbol{r}\right]_{E=E_0}\right)}{\delta E + \frac{i\Gamma}{2}} \tag{15}$$

with respect to $\delta E$. The derivative in Eq. (14) is easy to evaluate. From Eq. (14) we have

$$1 = \frac{\partial E_F\left(\frac{\boldsymbol{r}}{\tau}\right)}{\partial E} = -\sum_n \frac{\partial E_F\left(\boldsymbol{v} = \frac{\boldsymbol{r}}{\tau}\right)}{\partial v_n} \frac{r_n}{\tau^2} \frac{\partial \tau}{\partial E}. \tag{16}$$

This means (suppressing the subscript "$s$")

$$\frac{\partial}{\partial E}(\boldsymbol{p}\cdot\boldsymbol{r}) = -\sum_{n,m} r_n \frac{\partial p_n}{\partial v_m} \frac{r_m}{\tau^2} \frac{\partial \tau}{\partial E} = -\sum_{n,m} \frac{\partial E_F}{\partial p_n} \frac{\partial p_n}{\partial v_m} \frac{r_m}{\tau} \frac{\partial \tau}{\partial E} = -\sum_{n,m} \frac{\partial E_F}{\partial v_m} \frac{r_m}{\tau} \frac{\partial \tau}{\partial E} = \tau \quad (17)$$

Finally, the $\delta E$ integrand for the wavefront is

$$\frac{e^{-i(E_0+\delta E)t} e^{[i\boldsymbol{p}_s\cdot\boldsymbol{r}]_{E=E_0}} e^{i\tau_0 \delta E}}{\delta E + \frac{i\Gamma}{2}}, \quad (18)$$

where $\tau_0 = \tau(\boldsymbol{r}, E=E_0)$. Evaluating the integral, Eq (18) gives the following time-dependence, ignoring time-independent phases and other factors:

$$e^{-i\left(E_0-\frac{i\Gamma}{2}\right)(t-\tau_0)} \theta(t-\tau_0). \quad (19)$$

This generalization of Equation (3) is our first major result. In particular, we learn that the wavefront (leading edge) is always a surface of constant $\tau$.

To make this more tangible, let us write closed-form expressions for $\tau$ in particular limits. In the nonrelativistic case, we have the explicit formula

$$\tau = \sqrt{\frac{\sum_n \frac{1}{2} m_n r_n^2}{E}}. \quad (20)$$

In the relativistic case, we have the implicit formula

$$E = \sum_n m_n \rho_n \equiv \sum_n \frac{m_n}{\sqrt{1-\left(\frac{r_n}{\tau}\right)^2}}. \quad (21)$$

Clearly a nonrelativistic wavefront is an ellipsoid in $\boldsymbol{r}$ space. If some particles are relativistic, i.e. $E \gg m_n$ for some $n$, then the wavefront is close to $r_n = \tau$ for at least one such relativistic $n$, or all such $n$ are unconstrained, and the remaining radii are constrained to a nonrelativistic ellipsoid.

It will be useful in the next section to have noted here that Equation (14) implies that $t$ is homogeneous of degree one as a function of $\boldsymbol{r}$. The same is clearly also true of $\boldsymbol{r}\cdot\boldsymbol{p}_s$.

## 4. Pseudo-norm for multi-particle Gamow states

In order to generalize Equations (4) and (5), let us begin by reviewing their original derivation [3], so we can understand what we're actually trying to generalize. The single-spinless-particle Green's function $G(\boldsymbol{x},\boldsymbol{x}',E)$ for outgoing waves is defined in terms of the Schroedinger Hamiltonian $H$ via

$$(E - H)G = \delta^3(x - x'), \tag{22}$$

with boundary condition

$$\nabla_x G(x, x', E) = i\hat{x}\sqrt{2mE}\, G(x, x', E), \tag{23}$$

in the limit of very large |**x**|. $H$ on the left of Eq. (22) acts on $G$'s **x** coordinate, and the square root in Eq. (23) has positive real part for positive $E$. Near the decay resonance at $E_D=E_0-i\Gamma/2$, Garcia-Calderon and Peierls write

$$G(x, x', E) \sim \frac{C(x, x')}{E - E_D} + \chi(x, x'), \tag{24}$$

Substituting Eq. (24) into Eqs. (22) and (23), and zeroing the poles, $C$ satisfies

$$(E_D - H)C(x, x') = 0 \tag{25}$$

with outgoing boundary condition at very large |**x**|

$$\hat{x} \cdot \nabla_x C(x, x') = ik_D C(x, x'), \tag{26}$$

where $k_D$ is defined in Equation (6). From the pole remainders, one also obtains

$$(E_D - H)\chi(x, x') + C(x, x') = \delta^3(x - x') \tag{27}$$

with outgoing boundary condition

$$(\hat{x} \cdot \nabla_x - ik_D)\chi(x, x') = -\frac{im}{k_D} C(x, x'), \tag{28}$$

again in the limit of very large |**x**|.

Ignoring degeneracy unrelated to global symmetry, equations (25) and (26) imply $C(\mathbf{x},\mathbf{x}')=u(\mathbf{x})P(\mathbf{x}')$ for some function $P$. Equations (25) and (27) imply

$$[\chi(x, x')Hu(x) - u(x)H\chi(x, x')] + u^2(x)P(x') = u(x)\delta^3(x - x'). \tag{29}$$

One recognizes the square brackets in Eq. (29) as the usual total derivative characteristic of the free Schroedinger equation, so we can integrate Eq. (29) over **x** within a sphere of large radius $R$ to get

$$\oint_{|x|=R} \hat{x} \cdot \left(\frac{1}{2m}\right)[u(x)\nabla_x\chi(x,x') - \chi(x,x')\nabla_x u(x)]\, d^2x + \left[\int\!\!\int_{Inside} u^2(x)\, d^3x\right]P(x')$$
$$= u(x'). \tag{30}$$

Using boundary conditions (26) and (28), and continuing to factor $C$, this becomes

$$-\frac{i}{2k_D}\left[\oint_{|x|=R} u^2(x)d^2x\right]P(x') + \left[\int\!\!\int_{Inside} u^2(x)\, d^3x\right]P(x') = u(x'), \tag{31}$$

from which Equations (4) and (5) follow immediately.

We can now proceed to generalize all this to $N>1$ particles. The obvious direct generalization of Equation (30) is

$$\oint_{Boundary}\left\{u(x)\left[\sum_n \hat{s}_n \cdot \left(\frac{1}{2m_n}\right)\nabla_{x_n}\right]\chi(x,x') - \chi(x,x')\left[\sum_n \hat{s}_n \cdot \left(\frac{1}{2m_n}\right)\nabla_{x_n}\right]u(x)\right\}d^{3N-1}x$$
$$+ \left[\int\!\!\int_{Inside} u^2(x)\, d^{3N}x\right]P(x') = u(x'). \tag{32}$$

where **s** is a $3N$-dimensional boundary-surface normal. [This is only true if the potential function is symmetric under interchange of the coordinates and internal indices of two different particles at a time.] The obvious generalization of boundary condition (23) follows directly from the asymptotic form $u \sim \exp(i\mathbf{p}_s \cdot \mathbf{r})$ and the homogeneity of $\mathbf{p}_s \cdot \mathbf{r}$:

$$\sum_n x_n \cdot \nabla_{x_n} G(x,x',E) = i(\mathbf{p}_s \cdot \mathbf{r})G(x,x',E). \tag{33}$$

This flows down to boundary condition (26) as

$$\sum_n x_n \cdot \nabla_{x_n} C(x,x') = i(\mathbf{p}_s \cdot \mathbf{r})_{E=E_D} C(x,x'). \tag{34}$$

Recalling Equation (17), Equation (33) also flows down to boundary condition (28) as

$$\left[\left(\sum_n x_n \cdot \nabla_{x_n}\right) - i(\mathbf{p}_s \cdot \mathbf{r})_{E=E_D}\right]\chi(x,x') = -i\tau|_{E=E_D} C(x,x'). \tag{35}$$

To complete the generalization, we must transform the gradient sums of Equations (34) and (35) so that factors $(1/m_n)$ appear, and so that the ensemble $\{x_n\}$ is replaced by the normal to a convenient boundary in $3N$-dimensional position space. Here is how to do this: Begin by recognizing that for outgoing states, at large distances, we can write

$$\sum_n x_n \cdot \nabla_{x_n} \sim \sum_n r_n \cdot \frac{\partial}{\partial r_n} = \mathbf{r} \cdot \nabla_r. \tag{36}$$

Then decompose

$$\mathbf{r} \cdot \nabla_r = \tau \sum_n \frac{\partial E_F}{\partial p_n} \frac{\partial}{\partial r_n} = \tau \sum_n \frac{\partial E_F}{\partial v_n} \sum_m \frac{\partial v_n}{\partial p_m} \frac{\partial}{\partial r_m}$$
$$= \tau \sum_n \frac{\partial E_F}{\partial v_n} \sum_m \frac{\partial^2 E_F}{\partial p_n \partial p_m} \frac{\partial}{\partial r_m} \equiv \tau \sum_n \frac{\partial E_F}{\partial v_n} \sum_m (M^{-1})_{nm} \frac{\partial}{\partial r_m}. \quad (37)$$

(all $p_n$ and $v_n$ differentiation is at the stationary phase point). Then transform the derivative with respect to $v_n$ by exploiting

$$0 = \frac{\partial E_F}{\partial r_n} = \sum_m \frac{\partial E_F}{\partial v_m} \left( \frac{1}{\tau} \delta_{mn} - \frac{r_m}{\tau^2} \frac{\partial \tau}{\partial r_n} \right) \quad (38)$$

to arrive at

$$\mathbf{r} \cdot \nabla_r = \left( \sum_k r_k \frac{\partial E_F}{\partial v_k} \right) \sum_n \frac{\partial \tau}{\partial r_n} \sum_m (M^{-1})_{nm} \frac{\partial}{\partial r_m}. \quad (39)$$

In the interest of aesthetics, we can go further and transform the multiplier on the right hand side of Equation (39) to get

$$\left( \sum_k r_k \frac{\partial E_F}{\partial v_k} \right) = \tau \left( \sum_k \frac{\partial E_F}{\partial p_k} \frac{\partial E_F}{\partial v_k} \right) = \tau \left( \sum_{kn} \frac{\partial E_F}{\partial p_k} \frac{\partial E_F}{\partial p_n} \frac{\partial p_n}{\partial v_k} \right) = \tau \sum_{kn} \frac{\partial E_F}{\partial p_k} M_{kn} \frac{\partial E_F}{\partial p_n}. \quad (40)$$

This leads, in an obvious notation, to
,
$$\mathbf{r} \cdot \nabla_r = \tau \langle \nabla_p E_F | M | \nabla_p E_F \rangle \langle \nabla_r \tau | M^{-1} | \nabla_r \rangle. \quad (41)$$

The last factor in Equation (41) is precisely the operator in square brackets in Equation (32), where the unit vector **s** points orthogonal to a surface of constant $\tau$, except for a missing multiplier

$$(2T)^{-1} \equiv \left( 2 \sqrt{\sum_n \left( \frac{\partial \tau}{\partial r_n} \right)^2} \right)^{-1}. \quad (42)$$

The quantity $T$ is not literally a norm, because $\tau$ can be complex through its dependence on $E$; but for nonrelativistic free dynamics, the $E$ dependence cancels in the ratio $T^{-1} \nabla t$.

Finally, using Equations (34), (35) and (41) we can rewrite Equation (32) as

$$\left[\oint_{\tau=\tau_R} \frac{-i}{\left[2T\langle\nabla_p E_F|M|\nabla_p E_F\rangle\right]_{E=E_D}} u^2(\mathbf{x}) d^{3N-1}x\right] P(\mathbf{x}') + \left[\int_{Inside} u^2(\mathbf{x}) \, d^{3N}x\right] P(\mathbf{x}')$$
$$= u(\mathbf{x}'), \quad (43)$$

in the limit of very large of $\tau_R$. This means the multi-particle generalization of the Gamow state pseudo-norm must be

$$\mathcal{N} = \left[\int_{Inside} u^2(\mathbf{x}) \, d^{3N}x\right] - \left[\oint_{\tau=\tau_R} \frac{i}{\left[2T\langle\nabla_p E_F|M|\nabla_p E_F\rangle\right]_{E=E_D}} u^2(\mathbf{x}) d^{3N-1}x\right]. \quad (44)$$

in the large-$\tau_R$ limit. For the particular case of the nonrelativistic energy in Equation (7), the denominator in the second term in Eq. (44) works out to be

$$\left[2T\langle\nabla_p E_F|M|\nabla_p E_F\rangle\right]_{E=E_D} = \frac{2}{\tau}\sqrt{\sum_n (m_n r_n)^2}, \quad (45)$$

where everything is evaluated on the surface $\tau=\tau_R$. Equations (44) and (45) together constitute our second major result. It obviously reduces to Equation (5) in the single-particle limit.

## 5. Concluding remarks

Single-particle Gamow states represent an interesting extension of the familiar mathematical formalism of normalizable bound states, and have found applications from nuclear shell models to the foundations of quantum mechanics. In this paper, we have shown that Gamow states and their main properties can be generalized to multi-particle scenarios. We have derived closed-form expressions for the wavefront of a multi-particle Gamow state, and for the pseudo-norm that sets the scale for such a state's contribution to a quantum Green's function. It is hoped that these new results can set the stage for further practical applications and fundamental insights.